\documentclass[journal=macro,manuscript=article]{achemso}

\usepackage{graphicx}
\usepackage{dcolumn}
\usepackage{bm}
\usepackage{svg}
\usepackage{amsmath}
\usepackage{mathtools}
\usepackage{placeins}
\usepackage{float}
\usepackage{url}
\usepackage{xcolor}
\usepackage{tikz}
\usepackage{amssymb}
\usepackage{caption}
\usepackage{booktabs}
\usepackage{subcaption}
\usepackage{upgreek}
\usepackage[T1]{fontenc}
\usepackage{hyperref}
\hypersetup{
    colorlinks=true,linkcolor=blue, urlcolor=gray, citecolor=blue}
\usepackage{romannum}
\AtBeginDocument{\pagenumbering{arabic}}
\usepackage[super]{nth}
\usepackage[newcommands]{ragged2e}
\usepackage{siunitx}
\DeclareUnicodeCharacter{2009}{\,}
\DeclareUnicodeCharacter{3B1}{\ensuremath{\alpha}}
\DeclareSIUnit\angstrom{\text {Å}}
\usetikzlibrary{plotmarks}
\author{Andrzej Grzyb}
\affiliation{Institute of Physics, University of Silesia in Katowice, 75 Pu\l{}ku Piechoty 1, 41-500 Chorz{\'o}w, Poland}
\author{Khristine Haydukivska}
\affiliation
{Institute of Physics, University of Silesia in Katowice, 75 Pu\l{}ku Piechoty 1, 41-500 Chorz{\'o}w, Poland}
\alsoaffiliation{Institute for Condensed
Matter Physics of the National Academy of Sciences of Ukraine,\\
79011 Lviv, Ukraine}
\author{Jaros{\l}aw S. K\l{}os}
\affiliation{Faculty of Physics, A. Mickiewicz University, Uniwersytetu Pozna{\'n}skiego 2, 61-614 Pozna\'n, Poland}
\alsoaffiliation{Leibniz-Institut f\"ur Polymerforschung Dresden e.V., 01069 Dresden, Germany}
\author{Aykut Erba\c{s}}
\affiliation{Institute of Physics, University of Silesia in Katowice,
75 Pu\l{}ku Piechoty 1, 41-500 Chorz{\'o}w, Poland}
\alsoaffiliation{UNAM - National Nanotechnology Research Center and Institute of Materials Science \& Nanotechnology, Bilkent University, 06800 Ankara, Turkey}

\author{Jaros{\l}aw Paturej}
\affiliation{Institute of Physics, University of Silesia in Katowice,
75 Pu\l{}ku Piechoty 1, 41-500 Chorz{\'o}w, Poland}
\alsoaffiliation{Leibniz-Institut f\"ur Polymerforschung Dresden e.V., 01069 Dresden, Germany}
\email{jaroslaw.paturej@us.edu.pl}

\title{Structural properties of\\ cyclic polyelectrolytes\\ in dilute good solvent}

\begin{document}




\clearpage
\begin{abstract}
\singlespacing

Cyclic polymers display unique physical behaviors in comparison to their linear counterparts. Theoretical, computational and experimental studies have revealed that some of their distinctive properties are also observed in charged variants of cyclic polymers, known as cyclic polyelectrolytes (PEs), especially in terms of their structural responses to variations in the strength of electrostatic interactions. In this study, we investigate the impact of cyclic topology on the conformations of PE chains in dilute good solvent using scaling analysis and coarse-grained bead-spring molecular dynamics simulations.
Our observations indicate that, in contrast to linear PE chains, cyclic topology results in  more compact conformations at low and intermediate Bjerrum lengths.   
Moreover, two structural metrics, asphericity and prolateness, which quantify deviations from spherical and flat molecular shapes, exhibit non-monotonic behaviors for cyclic PEs. 
This stands in contrast to linear PEs, where these shape characteristics exhibit a monotonic trend with increasing Bjerrum length.
A feasible analytical theory, developed to account for ionic distributions around cyclic PE chains, suggests that the fundamental difference between linear and cyclic chain conformations may be attributed to topological effects influencing long-range electrostatic interactions.

\end{abstract}
\clearpage

\maketitle


\section{Introduction}
\hspace*{.5cm}Polyelectrolytes (PEs) are polymers that contain ionizable functional groups  along their backbones.
When PE chains are dissolved in polar solvents,  ionizable groups can release counterions into solution \cite{hara}, leaving a net charge on the polymer backbone.
Mobile counterions make PEs solutions highly responsive to changes in pH, ionic strength, and electric fields \cite{muthu_review}. For the same reason, while synthetic PEs are commonly found throughout various applications 
 including industrial water treatment \cite{bolto}, dispersant agents, emulsifiers \cite{kogej}, superplasticizers \cite{superplast}, gene therapy and drug coating/delivery \cite{pack}, tissue engineering and cosmetics,
biological PEs such  DNA plasmids play a dominant role in many biochemical processes \cite{manning,achazi}.

PE chains share properties with both electrolytes and polymers, and the interplay of these properties determines their equilibrium configuration. 
Electrostatic interactions, counterion entropy, and conformational free energy of a polymer chain  affect the degree of counterion delocalization from polymer backbone \cite{barrat}. Ions localized near or on the chain, on the other hand, can neutralize the backbone charges, and in turn, change the equilibrium structure of the PE chain~\cite{mann,liu}. For instance, in linear PEs, the ionic strength of the solution controls the polymer-counterion interactions and can determine whether the chain  adapts stretched,  pearl-necklace or collapsed conformations~\cite{dobrynin1,dobrynin2}.
The coupling between PE counterions and chain conformation can further be affected by charge distribution along the polymer backbone, charge valence, polymer chemistry, degree of polymerization, and macromolecular architecture.

Introducing structural complexity in a macromolecular architecture, for instance, in the form of grafts, branching points, and closed loops, can expand the possibilities for designing novel polymers with distinctive physiochemical properties that their linear counterparts cannot achieve.
Cyclic polymers are an intriguing class of such polymers with loop topology. In  solutions of charge-neutral cyclic polymers, lack of ends enforces a smaller radius of gyration, smaller hydrodynamic volume, lower melt viscosity, higher thermostability, and a higher glass transition temperature as compared to linear chains \cite{zimm,roovers,laurent,halverson,tezuka,jia,zhang,lang,hossain}.
The compactness of cyclic polymers plays a crucial role in controlling the segregation properties of colloidal particles in confinement \cite{erbas_colloid}. Additionally,   cyclic topology strongly influences specific viscoelastic properties of polymer melts, such as the absence of plateau modulus that is common in melts of entangled linear chains \cite{lang}. Cyclic polymers also exhibit significantly larger swelling ability and maximum strain at break in their swollen networks \cite{tew} and enhanced lubrication abilities\cite{erbas,morgese}.

Synthetic cyclic PEs, which are experimentally available through various synthetic strategies \cite{hiroaki,chen2018,jainhao}, have also shown peculiar behaviour similar to their neutral counterparts in experimental and theoretical studies.
Thermodynamics of ring PE solutions was investigated using analytical methods \cite{BERNARD2017121}. The authors found that at semidilute concentrations, the electrostatic contribution to osmotic pressure is larger for PE rings as compared to linear chains.  
Molecular dynamics (MD) studies on counterion distribution around highly-charged cyclic, linear, and branched PE architectures  in polar solvent  demonstrated the effect of macromolecular complexity on the fraction of counterions transiently localized to PE chains~\cite{chremos}. 
 Linear and cyclic PE chains grafted on planar surfaces exhibit variation in surface morphologies in MD simulations~\cite{miao}. While both linear and cyclic polymers exhibit a formation of vertically phase-separated patched structures depending on the grafting density, cyclic topology favours smaller patches.
%
%
Monte-Carlo simulations \cite{nowicki} suggested that
charged ring-shaped macromolecules experience a significantly more pronounced segregation confined in  nanochannels.

Liu {\it et al.} studied the motion of circular PEs in shear flow by means of mesoscale hydrodynamic simulations and revealed
coupling between shear rate, electrostatic strength and polymer architecture and their impact on conformation and dynamics of charged rings \cite{an}.  The circular architecture plays a pivotal role in defining deformation and orientation characteristics of  charged rings subjected to flow. 
If the electrostatic interaction strength is weak (i.e.~at high salt concentration), the PE chain shifts its shape from  an oblate ring at low shear rates to  a prolate ring at high shear rates. Conversely, when the electrostatic interaction strength is strong (i.e.~at low salt concentration), the transition occurs from a prolate coil to a prolate ring.
Last but not least, 
biological cyclic PEs such as bacterial genome and plasmids somewhat prefer cyclic topology to optimize various vital biological processes such as loop formation, replication, recombination, repair, gene sequencing and coding \cite{loop1,loop2,replica,recomb,repair,gene,coding}.

In the present work, we study the impact of topology on conformations of PEs in salt-free good solvent conditions using analytical approaches and bead-spring coarse-grained MD simulations. We systematically investigate the role of polymerization degree and the Bjerrum length (i.e. strength of electrostatic interactions) on cyclic PE chains size, shape and monomer density and compared them to linear PEs of the same degree of polymerization. We observe significant differences in the equilibrium conformation properties of cyclic PEs, some of which can be described by our scaling arguments suggested at the strongly-stretched chain limit.

The paper is organized as follows. In the next section, we outline the details of the simulation model utilized in this study.  We apply scaling analysis to estimate conformations of cyclic PE chains.  Then, we present numerical results and discussed them in the context of our theoretical analysis. We draw conclusions and remarks in the final section of the manuscript.

\section{Simulation Model and Methodology}

Linear PEs are long-chain polymers with repeating units that contain charged functional groups. Ring polyelectrolytes are polymers with a closed-loop or circular architecture. Like linear polyelectrolytes, they contain repeating units with charged functional groups, which can be positively charged (cationic) or negatively charged (anionic).  In our study we utilize molecular dynamics (MD) simulations, implementing the well-established Kremer-Grest bead-spring model that was previously employed in investigations of linear PEs. \cite{mica,micka2,liao,dobrynin}. In the KG model, non-bonded interactions between all beads are accounted by a shifted, truncated LJ potential \cite{mica}:
\begin{equation}
U^{\text{LJ}}\left(r_{ij}\right)=
\begin{cases}
4\epsilon \left[\left(\frac{\sigma}{r_{ij}}\right)^{12}-\left(\frac{\sigma}{r_{ij}}\right)^{6} -\left(\frac{\sigma}{r_c}\right)^{12}
+\left(\frac{\sigma}{r_c}\right)^{6} \right] & \text{for }r \leq r_{c}  \\
0 & \text{for } r>r_{c} \\
\end{cases}
\label{eq:LJ}
\end{equation}
where $r_{ij}$ is the distance between two interacting beads $i$ and $j$, $\sigma$ is the bead diameter and $r_c$ is the cutoff distance. The parameter $\epsilon$ in Eq.~(\ref{eq:LJ}) controls the strength of  the LJ potential. The connectivity between two adjacent chain monomers is maintained by the finite extension nonlinear elastic (FENE)
potential
\begin{equation}
U^{\text{FENE}}\left(r\right)=
\begin{cases}
-0.5kr_{\text{max}}^2 \ln {\left(1-\frac{r^2}{r_{\text{max}}^2} \right)}, & \text {if } r \leq r_{\text{max}} \\
\infty, & \text{if } r > r_{\text{max}}\\
\end{cases}
\label{eq:FENE}
\end{equation}
where $k$ is the spring constant and $r_{\text{max}}$ is the maximum bond length \cite{mica}. To investigate linear and ring PEs monovalent positive charges are assigned to each bead along the polymer backbone. Furthermore, an appropriate number of monovalent, negative counterions are considered as well to maintain system electroneutrality. The electrostatic interactions  between charges were introduced by Coulomb pairwise potential
\begin{equation}
\frac{U^{\mbox{\tiny C}}_{ij}}{k_BT}=l_B\frac{z_{i}z_{j}}{r}
\label{eq:coulomb}
\end{equation}
where $k_{B}$ denotes the Boltzmann constant, $T$ the absolute temperature,  $l_B$ the Bjerrum length of the solvent,
{{and $z_{i}$, $z_{j}=\pm 1$ is the charge valence of $i$-th and $j$-th particles.}}
{{The Bjerrum length is defined as $l_B={e^2}/({4\pi \varepsilon_0 \varepsilon_{r} k_BT})$,
where $\varepsilon_0$ stands for the electric permittivity of the vacuum, $\varepsilon_{r}$ for the relative permittivity of the solvent and $e$ for elementary charge. For instance, in water at room temperature, $T_r\approx 298$~K, the Bjerrum length is $l_B(\varepsilon_{r}=80,T=T_r)\approx 7${\AA}.}}

The LAMMPS MD package \cite{LAMMPS} was used to carry out Langevin dynamics in NVT ensemble in the reduced  LJ units. Namely, $\epsilon=1k_BT_r$ and $\sigma=l_B(\varepsilon_{r}=80,T=T_r)$ were taken as the real units of energy and length respectively. For the FENE potential the reduced parameters were: $k^*=k\sigma^2/\epsilon=7$ and $R_0^*=2$.\cite{mica} According to the formula for reduced elementary charge, $e^*=e/\left(4\pi \varepsilon_0 \epsilon \sigma \right)^{1/2}$, and for the reduced temperature, $T^*=k_BT/\epsilon$, we assigned them values $e^*=9$ and $T^*=1$. The size $\sigma^*=1$ and monomeric mass $m^* =1$ were chosen to be the same for all monomers and counterions. The cutoff distance $r_c^*$ in Eq.~(\ref{eq:LJ}) for monomer-monomer interactions was equal to $2.5$ and for monomer-counterion and counterion-counterion interactions $2^{1/6}$. Furthermore, for monomer-counterion and counterion-counterion interactions the LJ interaction strength parameter  
was set to $\epsilon^*=1$. Our simulations mimic good solvent conditions corresponding to the choice of $\epsilon^*=0.1$ for monomer-monomer interactions \cite{dobrynin}. Note that the value of $\epsilon^*\approx 0.5$ between monomers yields nearly $\theta$ conditions for charged polymers \cite{dobrynin,mica}. Coulomb interactions were calculated via Particle-Particle-Particle Mesh (PPPM) Ewald method with the error tolerance for force $10^{-5}$ and the real space cutoff distance $r_e^*= 30$.
The dumping parameter in the Langevin equation of motion was taken as $\gamma^*=\gamma/\tau=1$, where $\tau=\sigma\left(m/\epsilon \right)^{1/2}$ is the LJ time unit and $m$ is the assumed real mass unit. The reduced time step was $\delta t^*=\delta t/\tau = 0.005$.

Simulations were carried out in cubic boxes with periodic boundary conditions.
The simulation box was varied depending on the degree of polymerization $N$ of PEs to obtain an equal concentration of cationic and anionic monomers for all systems $\rho_{+}=\rho_{-}\approx 2.35\cdot 10^{-6}\sigma^{-3}$.
Throughout our study, we varied $l_B$ by adjusting the relative permittivity $\varepsilon_r$ of the medium. 
{ To achieve adequate equilibration of PE polymers, a sufficiently large number of MD steps was selected to facilitate the relaxation of the mean-square size to its equilibrium values. This criterion resulted in a range of simulation runs varying between $2\cdot10^7$ and $2\cdot10^9$ MD steps, depending on the length of the polymer chain.}
The simulation snapshots were rendered using the Visual Molecular
Dynamics program \cite{vmd}.  

\section{Results and Discussion}

\subsection{Theoretical considerations}

In the subsequent analysis, we thoroughly examine the conformations of linear and cyclic PEs by employing an analytical approach. We start the discussion by re-introducing
scaling arguments that are able to characterize the conformations of linear PEs \cite{deGennes76,gennes_book,pfeuty,khokhlov1982theory,dobrynin1995}, and we then extend these analyses to cyclic polymers for the first time. Our analysis will be restricted to  a weak-to-intermediate scale of Coulomb couplings (i.e., $l_B<7$)  for which chains are stretched. This is partially because at higher Bjerrum lengths, ionic correlations become more dominant, preventing the use of conventional scaling arguments. From a broad perspective, the scaling analysis
relies on the assumption of distinct length scales and incorporates the notion of an electrostatic blob \cite{deGennes76,gennes_book}. All the chemical details of chains are described by Kuhn segment size, $b$, the number of Kuhn segments $N$, and the fraction of charged monomers $f$.

\subsubsection{Linear PE chains}

In dilute solutions of neutral linear polymers, the average size of a chain  (i.e., end-to-end distance) is given by $R_e^{\rm n} \approx bN^{\nu}$,
where $\nu$ is the scaling exponent
which takes values of $\nu\approx 3/5$ for good solvent, $\nu =1/2$ for $\theta$ solvent, and $\nu =1/3$ for poor solvent. 
In  salt-free and dilute solutions of PEs at low-to-intermediate $l_B$ values, the Debye screening length is much larger than the average distance between individual molecules. In this case, intrachain electrostatic interactions dominate, outweighing the interchain interactions. The backbone charges on the chains  interact through unscreened long-ranged electrostatic repulsive forces, which results in a elongated chain size $R^0_e$, such that $R^0_e \approx b N $.     

In its simplest form, the conformation of a linear, elongated PE chain can be conceptualized as
a linear string of equally sized electrostatic blobs \cite{deGennes76}. 
The average size of a blob $D_e^0$ represents the smallest correlation length of electrostatic interactions and can be estimated using scaling arguments. 
Blobs are subsections of chains consisting of  $g_e^0$ monomers that remain unperturbed by electrostatic interactions, and their statistical properties resemble those of neutral chains in good solvents. Hence, the blob size $D_e^0$ is related to the number
$g_e^0$ of monomers within a blob through the equation: 
$D^0_e=b(g_e^0)^{\nu}$.
The electrostatic interactions  of monomers inside each blob are approximately of the order of thermal energy $k_BT$: 
$F_e=(qfg^0_e)^2/(4\pi\epsilon D^0_e)\approx k_BT$\cite{deGennes76}. By utilizing the definition of $l_B$, the latter equation yields:
$(fg^0_e)^2l_B/D^0_e\approx 1$ which leads to the following relationship $D^0_e=l_B(fg_e^0)^2$.  We can rewrite equations for $g^0_e$ and $D^0_e$ as:
\begin{equation}
 g^0_e=B^{\frac{1}{\nu-1}},  
\label{ge0}
\end{equation}
and
\begin{equation}
  D^0_e=bB^{\frac{\nu}{\nu-1}}.  
\label{De0}
\end{equation}
In Eqs.~(\ref{ge0})-(\ref{De0}), we introduced  a constant $B$ which depends on chain statistics and is given by
\begin{equation}
  B = (uf^2)^{\frac{\nu-1}{\nu-2}} = 
    \begin{cases}
      (uf^2)^{1/3} & \text{for $\nu=1/2$}\\
      (uf^2)^{2/7} & \text{for $\nu=3/5$},
      \label{B}
    \end{cases}       
\end{equation}
where $u\equiv l_B/b$.

At length scales larger than $D^0_e$, electrostatic interactions become dominant. As a result, the electrostatic blobs repel each other, leading to the formation of an extended chain consisting of electrostatic blobs of length $R_e^0$.
In this manner, the end-to-end distance $R^0_e$ of a PE chain is imposed by the geometry of a linear array of blobs\cite{deGennes76,pfeuty,dobrynin1995}:
\begin{equation}
R^0_e = \frac{N}{g^0_e}D^0_e  = bBN.
\label{R0_e}
\end{equation}
The above equation  reveals that while the PE chain exhibits an extended overall configuration, within the electrostatic blobs, the chain is coiled.   
The scaling model presented above is limited in that it assumes a constant tension along the chain, which is inaccurate due to the variation in electrostatic potential along the chain. Specifically, the electrostatic potential is higher at the center of the chain compared to the chain ends. To address this limitation, a more refined model\cite{joanny,liao} that incorporates the non-uniform stretching of a PE chain is discussed below. 

The charge distribution along a linear PE chain is an end effect, where the charges are situated at the minima of the electrostatic potential.
The reduced electrostatic potential $\phi(z)$  at position $z$ along the deformation axis of a PE 
chain is defined as a sum of all contributing charges along the chain, $\phi(z)=l_B\sum_iq_i/|z\mathbf{\hat e}_z -\mathbf{r}_i|$, where $\mathbf{\hat e}_z$ is a unit vector pointing along the deformation axis.  
In the case of an infinitely long chain, the average potential remains constant throughout the chain. However, for a finite chain, the charges at the ends have fewer neighbours compared to the center, resulting in inhomogeneous in the electrostatic potential \cite{joanny}. As a consequence, the nonuniform stretching of the linear chain occurs, with the central segments of the chain experiencing stronger electrostatic repulsion leading to their more significant stretching compared to the sections near the chain ends \cite{liao,joanny}.
The conformation of a linear PE chain is characterized by a trumpet-like shape and can be visualized as a linear array of electrostatic blobs 
with varying sizes~\cite{joanny}. The size $D_e(z)$ of blobs increases from the smallest blob in the middle of the chain towards the chain ends, where the $D_e(z)$ is the largest\cite{joanny,liao}. The relation between the size $D_e(z)$ and the number $g_e(z)$ of monomers per blob is: 
\begin{equation}
D_e(z)\approx bg^{\nu}_e(z).
\label{blob}
\end{equation}
In Refs.~\cite{joanny,liao}, the analysis focused on the conformations of linear PEs in a $\theta$ solvent (i.e., $\nu =1/2$). 
In the following, we modify the calculations of this analysis to implement them for good-solvent conditions (i.e., $\nu =3/5$), for which excluded-volume interactions become more pronounced. To do so, we effectively consider excluded-volume effects in our calculations by utilizing the scaling relation given in  Eq.~(\ref{blob}).   
The reduced potential $\phi(z)$ can be estimated from one-dimensional charge distribution as follows~\cite{joanny,liao}:
\begin{equation}
\phi(z) \approx fl_B\left( \int_{-R_e/2}^{z-D_e(z)/2} \frac{g_e(z')}{D_e(z')} \frac{dz'}{z-z'}
+\int^{R_e/2}_{z+D_e(z)/2} \frac{g_e(z')}{D_e(z')} \frac{dz'}{z'-z} 
+ \frac{g_e(z)}{D_e(z)}
\right).
\label{phi_def}
\end{equation}
By using Eq.~(\ref{blob}), the ratio $g_e(z)/D_e(z)$ in Eq.~(\ref{phi_def}) can be rewritten as
\begin{equation}
 \frac{g_e(z)}{D_e(z)}=
 \frac 1b\left(\frac{D_e(z)}{b}\right)^{\frac{1-\nu}{\nu}}.
\label{ratio}
\end{equation}

By assuming a weak variation of the blob size $D_e(z)$ along the stretching direction of the chain  and utilizing the relationship given in Eq.~(\ref{ratio}), we can simplify Eq.~(\ref{phi_def}) as 
\begin{equation}
\phi(z) \approx uf\left(\frac{D_e(z)}{b}\right)^{\frac{1-\nu}{\nu}}
\left( \ln{\left[ \frac{R_e^2-4z^2}{D_e^2(z)}  \right]+1}    \right).
\label{phi_integrated}
\end{equation}
The terms inside the logarithm impose that the above equation is only valid in the interval $-\frac{R_e-D^0_e}{2}<z<\frac{R_e-D^0_e}{2}$. 

The derivative of Eq.~(\ref{phi_integrated}) with respect to position provides an expression for the electrostatic force acting on each point of the chain along its most probable, strongly stretched conformation~\cite{joanny,liao}. This force stretches the segments inside the blob and is resisted by a blob tension $t(z)=D_e(z)/g_e(z)$. The balance between these forces determines the dependence of the electrostatic blob size on $z$ as
\begin{equation}
\frac 32 b^{\frac{2(1-\nu)}{\nu}} \frac{\rm {d}}{{\rm d}z} D_e^{\frac{2(\nu-1)}{\nu}}(z) 
\approx f \frac{{\rm {d}}\phi(z)}{{\rm {d}}z}. 
\label{FB_transformed}
\end{equation}
Solving Eq.~(\ref{FB_transformed}) 
leads to an expression for  $D_e(z)$ along the deformation direction 
\begin{equation}
D^{\rm L}_e(z) \approx D^{0}_e\left[  \ln{\left(\frac{R^2_e-4z^2}{2R_eD^{0}_e}\right)-1}  \right]^{\frac{\nu}{3(\nu-1)}},
    \label{blobL}
\end{equation}
where $D^{0}_e=b(uf^2)^{\nu/(\nu-2)}$ denotes the electrostatic blobs at chain ends. Eq.~(\ref{blobL}) indicates that the smallest electrostatic blob is located in the middle of the chain.  

The mean chain size $R_e$ can be obtained from the conservation condition of monomers \cite{joanny,liao}
\begin{equation}
    N = 2\int_0^{\frac{R_e}{2}-\frac{D^{0}_e}{2}}\!\!\rho(z){\rm{d}}z \approx \frac{R_e(D^{0}_e)^{\frac{1-\nu}{\nu}}}{b^{1/\nu}}
    \left( \ln{\frac{eR_e}{D^{0}_e}} \right)^{-1/3},
\label{Nlin}
\end{equation}
 where $\rho(z) = g_e(z)/D_e(z) $ is the projection of the monomer density onto the direction of the end-to-end vector and can be written using Eqs.~(\ref{ratio}) and (\ref{blobL}) as
\begin{equation}
\rho(z) \approx \rho_0[\ln{(R_e^2/4-z^2)+C}]^{-1/3},
\label{rhoz}    
\end{equation}
In Eq.~(\ref{rhoz}),  parameters $\rho_0$ and $C$  are  the monomer density away from the ends and  a constant, respectively. The expression provided in Eq.~(\ref{rhoz}) will be employed in our subsequent analysis to facilitate a comparison between scaling predictions and the numerical results obtained through MD simulations (see Fig.~\ref{fig:rho}a). 

By employing  Eq.~(\ref{rhoz}) and following  an iterative approach to solving Eq.~(\ref{Nlin}) 
 leads to an expression for the equilibrium size of a linear PE chains~\cite{liao,joanny}:
\begin{equation}
    R_e \approx b^{1/\nu} (D_e^0)^{\frac{\nu-1}{\nu}}[\ln{eN/g_e^0}]^{1/3}= bBN[\ln{eN/g_e^0}]^{1/3}.    
    \label{Re}
\end{equation}
The logarithmic term in Eq.~(\ref{Re}) arises due to the non-uniform stretching of linear PE chains. This logarithmic correction term is not observed in a simple equi-sized blob model (in contrast to Eq.~(\ref{R0_e})). Note that, similarly to Eq.~(\ref{R0_e}), the solvent quality effects are controlled by the prefactor $B$ defined in Eq.~(\ref{B}). For $\theta$-solvent condition with $\nu=1/2$ 
the scaling law of Eq.~(\ref{Re}) reduces 
to $R_e\approx bN(uf^2)^{1/3}[\ln{eN/g_e^0}]^{1/3}$ which is the result known from previous studies on linear PEs and was derived using force balance equation\cite{joanny,liao} and 
several other calculation techniques \cite{deGennes76,barrat2,vilgis}.  
For good solvent  with $\nu=3/5$, Eq.~(\ref{Re}) becomes
 $R_e\approx bN(uf^2)^{2/7}[\ln{eN/g_e^0}]^{1/3}$. Note that the logarithmic correction is not affected by the solvent quality.

\subsubsection{Cyclic PE chains}
\begin{figure}
    \centering
    \includegraphics[scale=0.2]{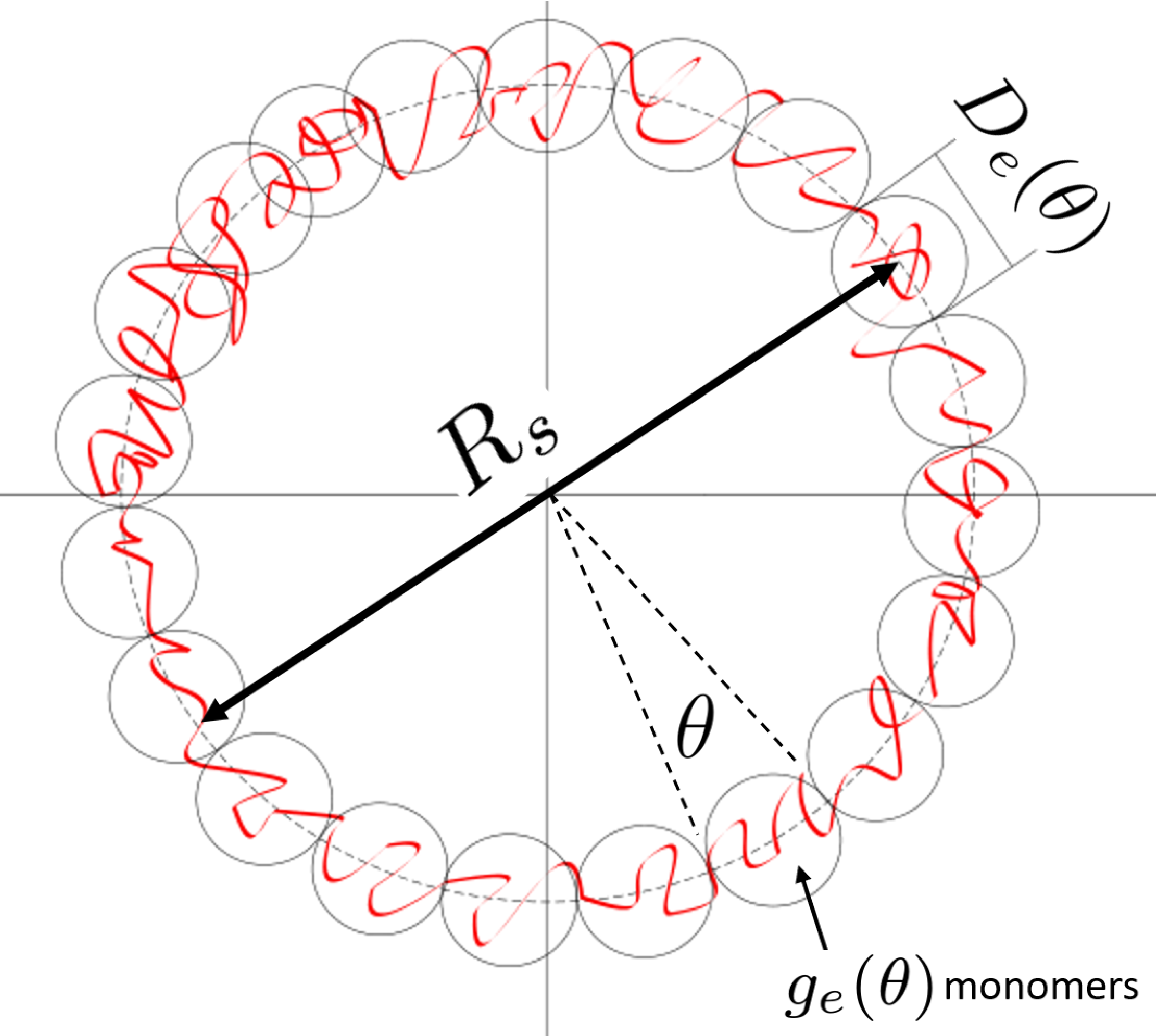}
    \caption{Schematic representation of a cyclic PE chain in a stretched conformational regime observed at low values of Bjerrum length $l_B/\sigma <7$. $R_s$ is the spanning { distance} of a ring whereas $D_e(\theta)$ and $g_e(\theta)$ denote the size of electrostatic blob and the number of monomers per blob, respectively.}
    \label{ring}
\end{figure}

If the two ends of a linear chain are attached covalently, the resulting ring polymer structure does not suffer from the non-uniform stretching  discussed in the previous section.  In this section, the constitutive equations we employ for linear chains will be modified to describe the equilibrium conformational properties of cyclic PE chains.  Fig.~\ref{ring} illustrates the schematics of a cyclic PE chain in its extended configuration, represented as a circular array of electrostatic blobs.  Due to the circular symmetry, the blob positions on the cyclic chain are described in terms of a polar angle $\theta$.  

%

Similar to Eq.~(\ref{phi_def}), the electrostatic potential of circular charge distribution can be estimated from
\begin{eqnarray}
&&\phi(\theta) 	\approx f l_B \left(\int_0^{\theta-\frac{D_e}{2}}d\theta' \frac{g_e(\theta')}{D_e(\theta')}\frac{\sqrt{2}}{R_s\sqrt(1-\cos{[\theta-\theta']})}\right.\nonumber\\
&&\left.+\int_{\theta+\frac{D_e}{2}}^{2\pi}d\theta' \frac{g_e(\theta')}{D_e(\theta')}\frac{\sqrt{2}}{R_s\sqrt(1-\cos{[\theta'-\theta]})}+\frac{g_e(\theta)}{D_e(\theta)}\right),
\label{phi_theta}
\end{eqnarray}
where $R_s$ { is the spanning distance} defined as the root mean-square  distance between monomers $1$ and $N/2 + 1$. 
Upon assuming a weak variation of the blob size $D_e(\theta)$ with the angle $\theta$, and making use of  Eq.~(\ref{ratio}) with angular dependence, we simplify Eq.~(\ref{phi_theta}) to the following form
{ 
\begin{equation}
   \phi(\theta) 	\approx uf^2
   \left(\frac{D_e(\theta)}{b}\right)^{\frac{1-\nu}{\nu}}\left(\frac{2}{R_s}\ln\left(\frac{1+\cos{[D_e^0/4]}}{1-\cos{[D_e^0/4]}}\right)+1\right).
\end{equation}
}
In order to obtain the electrostatic blob size $D_e(\theta)$, a force balance similar to that resulting in Eq.~(\ref{FB_transformed}) can be written as follows
\begin{equation}
   \frac 32 b^{\frac{2(1-\nu)}{\nu}}\frac{d}{d\theta}D_e(\theta)^{\frac{2(\nu-1}{\nu})}	\approx f \frac{d\phi}{d\theta},
\label{FB_circular}
\end{equation}
which balances blob tension  (i.e., $t(\theta) = D_e(\theta)/g_e(\theta)$) with electrostatic force acting on each point of the chain along its most probable strongly stretched circular contour.
The solution to Eq.~(\ref{FB_circular}) reads
{ 
\begin{eqnarray}
    &&\frac{3}{2}b^{\frac{3(1-\nu)}{\nu}}D_e(\theta)^{\frac{3(\nu-1)}{\nu}}\approx  \frac32 b^{\frac{3(1-\nu)}{\nu}}\frac{(D^0_e)^{\frac{2(\nu-1)}{\nu}}}{(D_e(\theta))^{\frac{1-\nu}{\nu}}}\nonumber\\
    &&+uf^2\left(1-\left(\frac{D^0_e}{D_e(\theta)}\right)^{\frac{1-\nu}{\nu}}\right)\left(\frac{2}{R_s}\ln\left(\frac{1+\cos{[D_e^0/4]}}{1-\cos{[D_e^0/4]}}\right)+1\right),
    \label{FB_circular_long}
\end{eqnarray}
}
where $D^0_e=b(uf^2)^{\nu/(\nu-2)}$.  
The above equation can be further simplified by employing the strong stretching approximation, where $D^0_e/D_e(\theta)\approx 1$. In the following sections, MD data simulations will verify this assumption {\it a posteriori}.  This approximation renders the contribution of the second term on the right-hand side of Eq.~(\ref{FB_circular_long}) negligibly small and leads to 
\begin{equation}
    D_e(\theta)\approx D^0_e.
    \label{blob_cyclic}
\end{equation}
The above expression provides an approximately constant blob size along the circular contour.

The total number of monomers in the ring is related to $R_s$ and $D^0_e$ as
\begin{equation}
    N\approx
    \frac{R_s}{2}\int^{2\pi}_0 \rho(\theta) d\theta=\frac{\pi R_s(D_e^0)^{\frac{1-\nu}{\nu}}}{b^{1/\nu}}.
    \label{Ncyclic}
\end{equation}
In the above equation, the density $\rho(\theta)$ is given by 
\begin{equation}
\rho(\theta) = \frac{g_e(\theta)}{D_e(\theta)} \approx \frac{(D_e^0)^{\frac{1-\nu}{\nu}}}{b^{1/\nu}}\approx \frac N{\pi R_s}.
\label{rhotheta}
\end{equation}
From Eq.~(\ref{Ncyclic}),  the mean size of a cyclic PE chain is obtained as
\begin{equation}
    R_s \approx \pi b^{1/\nu}(D_e^0)^{\frac{\nu-1}{\nu}}N \approx bBN.
\label{Rs}
\end{equation}
Notably, the above scaling relationship takes the form of Eq.~(\ref{R0_e}), and the influence of solvent quality is captured by the prefactor $B$ as in Eq.~(\ref{B}).  Importantly, due to the equal blob sizes of cyclic chains,  Eq.~(\ref{Rs}) does not incorporate any logarithmic corrections to the scaling as in the case of linear PE chain (cf.~Eq.~(\ref{Re})).

\subsection{Comparison between theory and MD simulations}

 We performed MD simulations of PEs exploring two different molecular architectures: linear chains and cyclic structures.
 In simulations, we systematically change the number of monomers per PE chain (i.e., $N=124, 244, 304$, and $604$). Each system is studied under a range of Bjerrum lengths ($0.125 \leq l_B \leq 27$), encompassing both weak and strong electrostatic coupling to examine the equilibrium conformational behaviour of the PEs across these extremes. 

\subsubsection{Equilibrium size of linear and cyclic PEs}
\begin{figure}[!h]
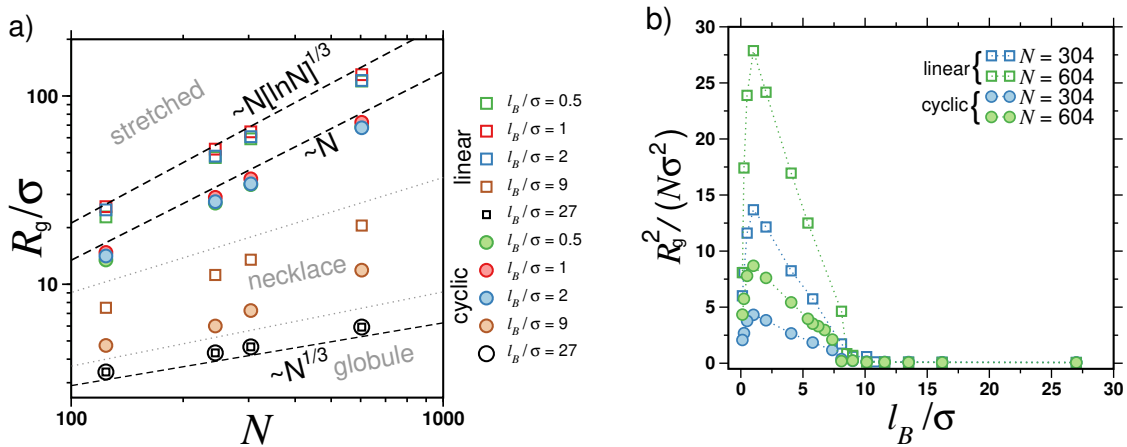

\includegraphics[scale=0.28]{sizes_v2.eps}
\hspace*{.5cm}
\includegraphics[scale=0.28]{rgVSlB_v2.eps}
\caption{\label{fig:size}
a) The { root mean-square} radius of gyration $R_g$ for linear (squares) and cyclic PEs for stretched, necklace and globule-like conformational regimes plotted as a function of degree of polymerization $N$ for different values of Bjerrum length $l_B$ as indicated. Dashed lines represent theoretical scaling laws in stretched ($R_g\propto  N(\ln{N})^{1/3}$ for linear and $R_g\propto N$ for cyclic architectures) and globule ($R_g\propto N^{1/3}$) conformations.
b) The variation of the rescaled mean-square radius of gyration $R_g^2/N$ of linear (squares) and cyclic (circles) PEs with $l_B$. The data are displayed for different values of $N$ as indicated.}
\end{figure}

\begin{figure}[!h]
\includegraphics[scale=0.8]{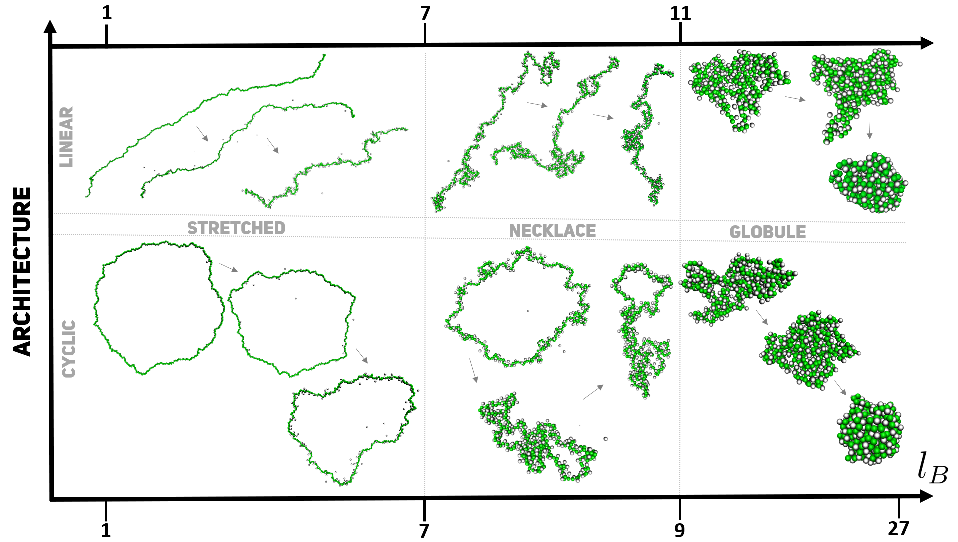}
\caption{\label{fig:snaps}
Molecular dynamics snapshots of linear (top panel) and cyclic (bottom panel) PEs displayed for different conformational regimes observed at different intervals of Bjerrum lenght $l_B$.
The degree of polymerization of all PEs is $N=304$.
 PE monomers are depicted as green beads, whereas
counterions are presented in silver.}
\end{figure}

We begin our analyses by focusing on the { root mean-square} radius of gyration $R_g$ of linear and cyclic PE chains.
In Fig.~\ref{fig:size}a, $R_g$ values calculated by averaging simulation trajectories are plotted as a function of polymerization degree $N$ for various  values of the Bjerrum length, $l_B$.  Regardless of  $l_B$, increasing $N$ leads to larger molecular sizes, but linear chains  are up to three times larger than cyclic chains except for the completely collapsed structures at $l_B / \sigma \gg 1$. For values of Bjerrum lengths of around $l_B / \sigma \lesssim 7$, both linear and cyclic chains adopt stretched conformations as can be seen also in  
Fig.~\ref{fig:snaps}, 
where we display the representative MD snapshots of PEs for both types of topologies.  In Fig.~\ref{fig:size}a, we also plot the scaling law predictions for chain sizes given in Eqs.~(\ref{Re}) and~(\ref{Rs}) in the stretched regime.  For a fair comparison, we first plot the scaling prediction for cyclic chains (i.e., $R_g \propto N$) and then multiply this prediction by the logarithmic correction term to describe the size of linear chains (i.e., $R_g \propto N[\ln{N}]^{1/3}$). The logarithmic correction terms, absent in the case of cyclic chains, serve to account for the more elongated configurations that linear chains can adopt.
At the strong electrostatic coupling limit (i.e., $l_B / \sigma \gg 1$), both linear and cyclic chains are collapsed (Fig.~\ref{fig:snaps}), and their sizes are statistically indistinguishable. In this regime, the chains dimensions are characterized by a scaling law of globular polymers $R_g \propto N^{1/3}$ \cite{dobrynin} (see also Fig.~\ref{fig:snaps}). 
In the intermediate range of electrostatic couplings (i.e., for $7\lesssim l_B / \sigma \lesssim 9$ ) the scaling exponent in the power law $R_g\propto N^{\nu_{\rm eff}}$  smoothly crosses-over from value $\nu_{\rm eff}=1$ to $1/3$.  In this regime both topologies adapt pearl-necklace conformations (cf.~Fig.~\ref{fig:snaps}).  

In Fig.~\ref{fig:size}b, the data given in Fig.~\ref{fig:size}a are plotted as a function of $l_B$ for two values of $N$. These data are in agreement with the results obtained in the previous numerical studies for linear PE chains\cite{liu,PhysRevLett.80.3731}.
Overall, we observe a non-monotonic behavior  of the chain sizes, regardless of topology. As $l_B \rightarrow 0$, the chains shrink in size due to weak electrostatic coupling and resulting weaker repulsion between backbone charges. In the stretched regime (i.e., for~$l_B / \sigma \approx 1$)
the majority of counterions are distanced from the chain, resulting in a negligible condensation effect and extended conformations for both types of architectures. 
A notable distinction between linear and cyclic PEs becomes evident in this regime. Linear PEs adopt one-dimensional rod-like conformations, whereas cyclic PEs take on a two-dimensional, ring-like structure that cover a broad surface area (see the snapshots in Fig.~\ref{fig:snaps}).
 For $l_B \rightarrow \infty$, the chains are strongly collapsed and take a globular conformation. In the globular regime, a great majority of the counterions condense on the chain (cf.~Fig.~\ref{fig:snaps}). For the intermediate values of $l_B$ (i.e., for $7\lesssim l_B / \sigma \lesssim 9$), an overall conformation is rather a combination of collapsed and stretched portions of the linear chains. Such conformations are reminiscent of the transition between the weak and strong coupling regimes and termed as pearl-necklace conformation in the literature~\cite{dobrynin,limbach,micka2,mica,rubinstein_obu}, (see also Fig.~\ref{fig:snaps}). 
This transition takes place as a result of the partial condensation of counterions onto the backbone, while the rest of the counterions become delocalized away from the chain. The transformation is also evident in Fig.~\ref{fig:size}a and b, where the fractal dimension characterizing the chain sizes undergoes changes in response to variations in the Bjerrum length. 
Hence, our simulations validate that such a transition takes place even when the chain backbone is in a good solvent condition.

While linear chains take a necklace conformation as predicted previously at the intermediate values of Bjerrum length~\cite{dobrynin}, (cf.~Fig.~\ref{fig:snaps}), for cyclic chains,  we also observe a qualitatively similar behavior. However, conformational fluctuations of cyclic chains are stronger than those observed for linear chains. While  we systematically and visually observe the necklace formation for linear chains, cyclic chains exhibit a transition between necklace to weakly collapsed conformations as can be seen in the snapshots in Fig.~\ref{fig:snaps}.  The collapsed states transiently emerge as multiple sub-globular sections approaching each other and forming a more crumbled chain conformation as compared to linear PEs. A visual inspection of simulation trajectories show that this transition is observed for all of our chain sizes and simulation replicas.

\begin{figure}[!h]
\includegraphics[scale=0.29]{pdf_size_ab_v2.eps}
\hspace*{.5cm}
\includegraphics[scale=0.665]{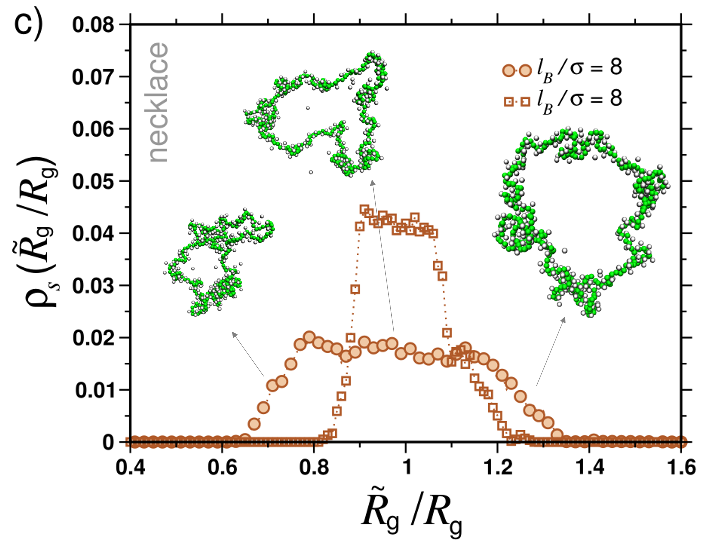}
\caption{\label{fig:pdfR}
Probability distribution functions $\rho_s$ of the radius of  gyration $\tilde{R}_g$ of linear (squares) and cyclic PEs normalized by its root mean-square value $R_g$
and plotted for different conformational regimes  observed at different values of Bjerrum lenght $l_B$. In panel data for a) stretched conformations ($l_B/\sigma=1$) are depicted. Panel b) shows data for globular conformations ($l_B/\sigma=27$), while panel c) data for  necklace conformations  { ($l_B/\sigma=8$)}. In panel c) arrows indicate MD snapshots of cyclic PEs. The middle snapshot shows conformation with size comparable to the average $\approx R_g$. The left and right snapshots correspond to the cases where molecular sizes are respectively smaller than $R_g$ (crumpled conformation) and larger than $R_g$ (expanded conformation). In all plots the degree of polymerization $N$ was fixed to $N=304$. 
} 
\end{figure}

To quantitatively characterize the conformational fluctuations of chains in different conformational regimes, we calculate probability distribution functions of { the instantaneous value of the radius of gyration $\tilde{R}_g$} for linear and ring chains for three values of Bjerrum length  (Fig.~\ref{fig:pdfR}). For both stretched and globule-like regimes,  the distribution functions are of a narrow width (Fig.~\ref{fig:pdfR}).  Nevertheless, cyclic chains  exhibit a slightly higher peak as compared to the linear chains. On the contrary, 
in the intermediate coupling regime where necklace structures are observed, there is an increase in the width of the distributions compared to the weak and strong coupling regimes for both topologies. However, cyclic chains { display a significantly broader distribution, suggesting larger conformational fluctuations and, thus, a more extensive spectrum of accessible conformations compared to their linear
counterparts.} 
This analysis is consistent with the visual observations made in the simulation snapshots displayed in Fig.~\ref{fig:snaps}.

\begin{figure}[!h]
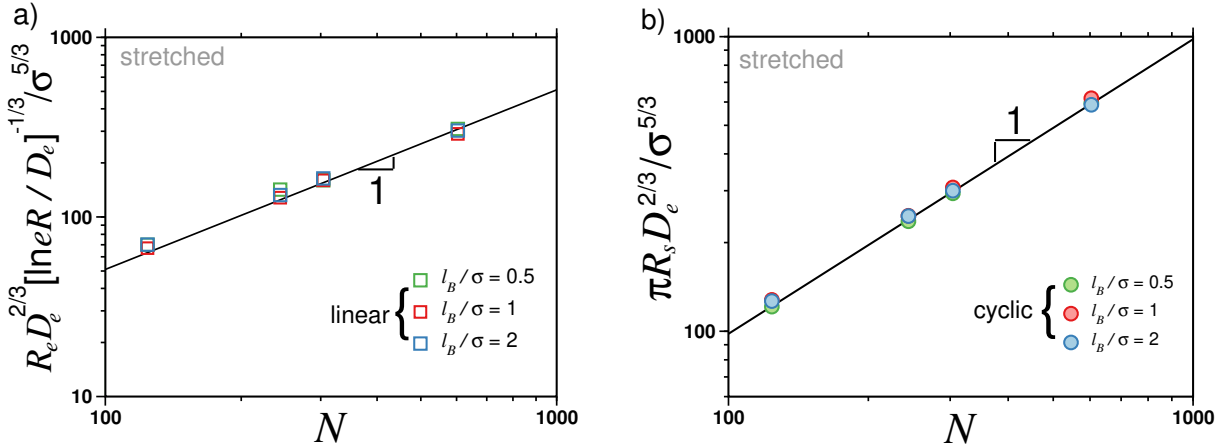

\includegraphics[scale=0.28]{size_scaling_linear_lowLb_v3.eps}
\hspace*{.5cm}
\includegraphics[scale=0.28]{size_scaling_ring_lowLb_v3.eps}
\caption{\label{fig:size_low}
The rescaled { root mean-square}: a) end-to-end distance $R_e$ for  linear PEs (squares) and b) spanning { distance} $R_s$ for cyclic  PEs (circles) plotted as a function of degree of polymerization $N$ for different values of Bjerrum length in the stretched regime as indicated. 
The lines represent theoretical scaling laws predicted for linear (Eq.~(\ref{Re})) and for cyclic (Eq.~(\ref{Rs})) architectures.
}
\end{figure}

In the previous sections, we described scaling model predictions for the mean size of linear (Eq.~(\ref{Re})) and cyclic (Eq.~(\ref{Rs})) PE chains at weak electrostatic couplings. 
In Fig.~\ref{fig:size_low}, we  compare the scaling predictions by considering the Bjerrum length $l_B / \sigma \leq 2$, corresponding to a regime, in which chains exist in stretched conformations. To construct the scaling plot, we calculate the average electrostatic blob size $D_e$ using the projected monomer density distributions  along the linear direction, $\rho(z)$,  and the angular direction, $\rho(\theta)$. The detailed analysis of monomer densities $\rho(z)$
and $\rho(\theta)$ are provided in the subsequent subsection. 
The obtained mean blob sizes for both types of architectures are listed in Table \ref{table}. We observe a strong concurrence between the theoretical predictions and the simulation results. The data for linear and cyclic chains collapse onto its own universal curve. Notably, the rescaled $y$-axes of the data for linear chains (Fig.~\ref{fig:size_low}a) indirectly validates the predicted non-uniform stretching along the chain, whereas the data for cyclic chains (Fig.~\ref{fig:size_low}b) indicates a uniform chain stretching. This observation highlights the distinct behavior of the two architectures at the weak electrostatic coupling regime.
Overall, our simulations show that cyclic PE chains exhibit a distinct conformational behavior, particularly from low to intermediate ionic strengths corresponding to  a Bjerrum length range of around $1<l_B / \sigma < 7$. 
In the next subsection, we characterize these differences by introducing further geometric  quantities characterizing the shape of PE in solution. 

\begin{table}[h!]
\begin{tabular}{cc|c|c}
 & & \multicolumn{2}{c}{$D_e/\sigma$} \\
\hline
 $N$ & $l_B/\sigma$   &linear & cyclic  \\
 \hline
124 & 0.5 &  1.84 & 1.58 \\
244 & 0.5 &  1.73 & 1.52  \\
304 & 0.5 &  1.60 & 1.50  \\
604 & 0.5 &  1.63 & 1.49  \\
124 & 1.0 &  1.73 & 1.49  \\
244 & 1.0 &  1.66 & 1.45  \\
304 & 1.0 &  1.67 & 1.44  \\
604 & 1.0 &  1.58 & 1.40  \\
124 & 2.0 &  1.54 & 1.59  \\
244 & 2.0 &  1.50 & 1.57  \\
304 & 2.0 &  1.51 &  1.52 \\
604 & 2.0 &  1.49 & 1.52\\
\caption{Mean values of electrostatic blob size $D_e$ for linear and cyclic PEs obtained for different values of degree of polymerization $N$ and Bjerrum length $l_B$. }
\label{table}
\end{tabular}
\end{table}


\subsubsection{Shape of linear and  cyclic PEs}

\begin{figure}[!h]
\includegraphics[scale=0.3]{shape_v2.eps}
\hspace*{.5cm}
\includegraphics[scale=0.7]{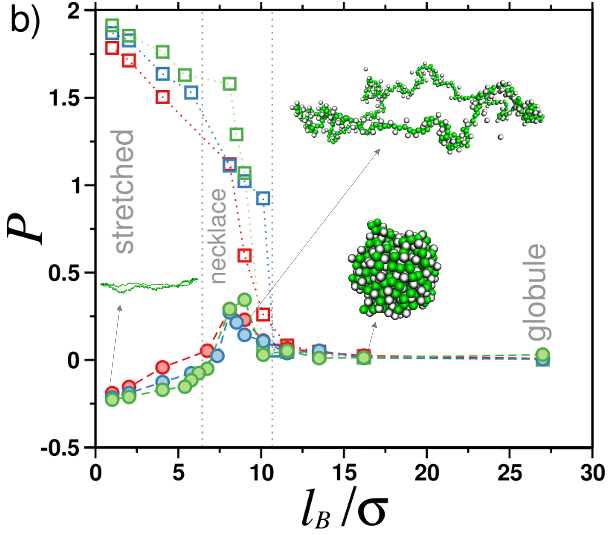}
\caption{\label{fig:shape}
a) Asphericity $A$  and b) prolatness $P$  shape factors for linear (squares) and cyclic (circles) PEs plotted as a function 
of the Bjerrum lenght $l_B$.
The data are displayed for different degrees of polymerization $N$ as indicated. 
In panel b), arrows are used to indicate molecular dynamics snapshots of cyclic molecules, displaying their characteristic shapes in different conformational regimes, namely the stretched, necklace, and globule-like conformations.
}
\end{figure}

Our simulations  show that cyclic PE chains exhibit distinct conformational fluctuations as the  Bjerrum length is varied (Fig.~\ref{fig:snaps}). These conformational fluctuations can affect the average shape of the molecules. Nevertheless, characteristic chain sizes fall short in analyzing such geometrical shape details. Alternatively, the average shape of a polymer chain  can be described by the average asphericity  $A$ and the average prolateness $P$ \cite{solc,theo,blavatska}. The quantity $A$ measures a degree to which a polymer conformation deviates from a perfect sphere. The parameter $P$ does  provide information about the flattening of a molecule~\cite{blavatska,aronovitz}.

The average asphericity can be defined as~\cite{solc,theo,blavatska}
\begin{equation}
 A=\frac{1}{6}\left\langle\sum_{i=1}^{3}\frac{ (\lambda_i - \bar{\lambda})^2}{\bar{\lambda}^2} \right\rangle,\label{Ad}
\end{equation}
where $\langle\ldots\rangle$ denotes ensemble average, $\lambda_i$  (for $i=1,..,3$) are the eigenvalues of the gyration tensor and  $\bar{\lambda}=(\lambda_1+\lambda_2+\lambda_3)/3$ is the average eigenvalue. 
The parameter $A$ is limited between the values of 0 
 and 1
and takes the minimum value for spherically symmetric 
configurations ($\lambda_1=\lambda_2=\lambda_3$) and the maximum value for rod-like configurations ($\lambda_1\neq 0$, $\lambda_2=\lambda_3=0$). For planar symmetric objects, 
$A$ converges to the value of $1/4$ \cite{theo}. 

The parameter $P$ is defined as\cite{blavatska,aronovitz}
\begin{equation}
 P=\left\langle\prod_{i=1}^{3}\frac{ (\lambda_i - \bar{\lambda})}{\bar{\lambda}^3} \right\rangle.\label{Pd}
\end{equation}
The values of $P$ are limited between the value of $-1/4$ for perfectly oblate, disk-like  shaped configurations 
($\lambda_1=\lambda_2$, $\lambda_3=0$) and the maximum value of $2$ for fully  prolate molecular configuration elongated  along one axis ($\lambda_1\neq 0$, $\lambda_2=\lambda_3=0$). In general, molecules with $P>0$ adapt prolate
 conformations ($\lambda_1\gg\lambda_2 \approx \lambda_3=0$), whereas molecules with $P<0$ adapt oblate conformations ($\lambda_1\approx\lambda_2 \gg \lambda_3=0$). The case of $P=0$ indicates
 an ideally spherical shape.  In other words, $P=0$ signals that the molecule is perfectly symmetrical in all directions, with no elongation or flattening along any axis.

In Fig.~\ref{fig:shape} we analyze the shape factors  of linear and cyclic PEs  by plotting $A$ and $P$  as a function of $l_B$ for chains with different $N$. 
It is found that an increase in electrostatic coupling induces a conformational transition in linear PEs, shifting them from a rod-like shape (with approximate values of $A\approx1$ and $P\approx2$) to a spherical shape (with $A\approx P\approx0$). Cyclic PEs exhibit distinct
conformations in the stretched and necklace regime observed up to $l_B/\sigma<9$. This is evident from the asphericity parameter $A\approx1/4$ which signals formation of planar (two-dimensional) configurations.
Furthermore, there is an observed increase in the $P$ parameter
with $l_B$ up to $l_B/\sigma<9$ suggesting a conformational transition in the shape of cyclic PEs (see snapshots in Fig.~\ref{fig:shape}b). In the stretched regime  cyclic PEs adapt nearly perfectly oblate (e.g., $P\approx -1/4$ for $l_B/\sigma=1$) shapes. In the necklace regime cyclic PEs become slightly prolate (e.g., $P\approx 1/4$ for $l_B/\sigma\approx 7)$ and exhibit wavy shape  due to formation of localized globular structures. Similarly to linear PEs, cyclic molecules tend to take a spherical shape  characterized by $A\approx P\approx0$ for larger values of $l_B$.  In summary, while shape transitions occur in both linear and cyclic chains as the Bjerrum length increases, cyclic chains exhibit unique, non-monotonic shape transitions, as clearly indicated by the asphericity and prolateness analyses (cf.~Fig.~\ref{fig:shape}). 


\subsubsection{Monomeric distributions and pair correlation functions  of linear and cyclic PEs}

\begin{figure}[!h]
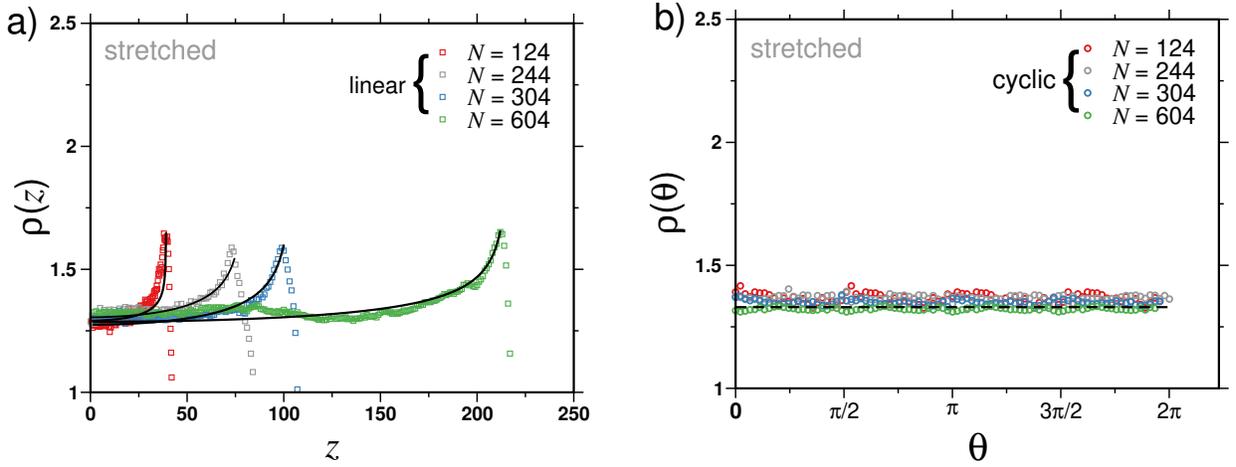

\includegraphics[scale=0.3]{rho.eps}
\hspace*{.5cm}
\includegraphics[scale=0.3]{rho_ring_v2.eps}
\caption{\label{fig:rho}
a) Projection of monomer density $\rho(z)$ of linear PEs on the direction of the end-to-end vector $R_e$ plotted from the position of a middle monomer. The solid lines are the best fit to Eq.~(\ref{rhoz}).
b) Angular monomer density $\rho(\theta)$ 
of cyclic PEs. The dashed line represents the density calculated from Eq.~(\ref{rhotheta})   
for $N=604$.   
In both panels the symbols are results of MD simulations for linear (squares) and cyclic (circles) PEs with different degree of polymerization $N$ as indicated. All the data are obtained for fixed Bjerrum lenght $l_B/\sigma =1$.}
\end{figure}


In the previous subsections, we present equations   for the projected and angular monomeric densities of linear and cyclic chains respectively (i.e., Eqs.~(\ref{rhoz}) and \ref{rhotheta}). 
Accordingly, while linear chains exhibit a non-uniform monomeric distribution due to non-uniform stretching, cyclic chains show a  rather uniform density profile. 
In Fig.~\ref{fig:rho}a and b, we compare these predictions with our MD results, namely  the projection of monomer density, $\rho(z)$, and  the angular density $\rho(\theta)$, for linear (squares) and cyclic (circles) PEs for various $N$. We restrict our analysis to the stretched regime (i.e. for $l_B/\sigma=1$), for which we have analytical equations. The corresponding theoretical predictions provided by Eqs.~(\ref{rhoz}) (solid lines) and (\ref{rhotheta}) (dashed line) are also shown in Fig.~\ref{fig:rho}. Overall, the fitting lines exhibit a high level of agreement with the MD data. The $\rho(z)$ profiles are in qualitative accordance with the findings of Ref.\cite{liao}, which studied linear PEs in $\theta$ solutions. In general, Fig.~\ref{fig:rho}a reveals that $\rho(z)$ increases with the distance $z$ from the chain center, indicating a nonuniform stretching of linear PEs towards their ends. Considering that the monomer density $\rho(z)$ is proportional to the electrostatic blob size $D_e(z)$ (cf.~Eq.~(\ref{rhoz})), the data in Fig.~\ref{fig:rho}a indicate the variation of $D_e(z)$ along the direction of chain stretching. In the case of cyclic PEs, the variation of blob size is absent. As displayed in Fig.~\ref{fig:rho} b) the monomer density $\rho(\theta)$ remains approximately constant along the circular contour, which suggests a uniform stretching of charged cyclic chains.

Next, we analyze the intrachain monomer-monomer correlation function  $g_{\rm {intra}}$(r), which is directly related to the probability of finding a pair of monomers on the same chain separated by the distance $r$. This quantity can further provide insights into internal polymer structure at length scales smaller than the chain size. The $g_{\rm {intra}}$ function is defined as follows:
\begin{equation}
    g_{\rm {intra}}(\vec{r})=\frac{1}{cN}\sum_{i\ne j} \langle \delta(\vec{r}-\vec{r}_{ij})\rangle
    \label{gintra_def}
\end{equation}
where 
$\langle \ldots \rangle$ denotes ensemble averaging.   To calculate $g^{\rm C}_{\rm {intra}}$ for cyclic PEs, we again consider their conformations  under the assumption of the strong stretching limit. Consequently, the fluctuations of monomer density along the circular contour are negligible in comparison to the average angular density $\rho(\theta)$. With this approximation, Eq.~(\ref{gintra_def}) can be reformulated as:
\begin{equation}
    g^{\rm{C}}_{\rm{intra}}(\vec{r})=\frac{1}{cN}\frac{R_s}{2}\int_0^{2\pi}\,d\theta\int_0^{2\pi}\,d\theta' \,\rho(\theta)\rho(\theta') \left\langle \delta\left(\vec{r}-R_s\sin\left(\frac{\theta-\theta'}{2}\right)\vec{e}\right)\right\rangle_{\rm o},
\label{gintra}
\end{equation}
where $\vec{e}$ is a unit vector along a line that connects two points on the rings and $\langle \ldots \rangle_{\rm o}$ represents averaging over all possible orientations of the vector.  As a result of this averaging  procedure, the following equation is obtained:
\begin{equation}
    g^{\rm{C}}_{\rm{intra}}(r)=\frac{R_s}{4\pi r^2 cN}\int_0^{2\pi}\,d\theta\,\rho(\theta)\rho(\theta+2\arcsin{ [r/R_s]})
    \label{gintra2}
\end{equation}
The expression given by Eq.~(\ref{gintra2})  can be further simplified  using constant monomer density $\rho(\theta)=N/(\pi R_s)$ along the ring:
\begin{equation}
    g^{\rm C}_{\rm{intra}}(r)=\frac{N(2\pi-2\arcsin{ [r/R_s]})}{2\pi^2r^2 c R_s}\theta(2\pi - 2\arcsin{ [r/R_s]}),
    \label{gC}
\end{equation}
where $\theta(x)$ is the Heaviside step function with the following properties: $\theta(x)=1$ for $x\geq 0$ and $\theta(x)=0$ for $x<0$. 
We note that the correlation function $g^{\rm L}_{\rm {intra}}$ for linear PEs in $\theta$ solvent was calculated using the strong limit approximation  in Ref.~\cite{liao}, and it takes the following analytical form:
\begin{equation}
   g^{\rm L}_{\rm{intra}}(r) = \frac{1}{2\pi c N r^2} \int^{R_e/2}_{-R_e/2}dz \rho(z) \rho(z+r) 
=   \frac {N}{2\pi c r^2}
   \frac{R_e - r}{R^2_e}\theta(R_e - r).
   \label{gL}
\end{equation}

\begin{figure}[!h]
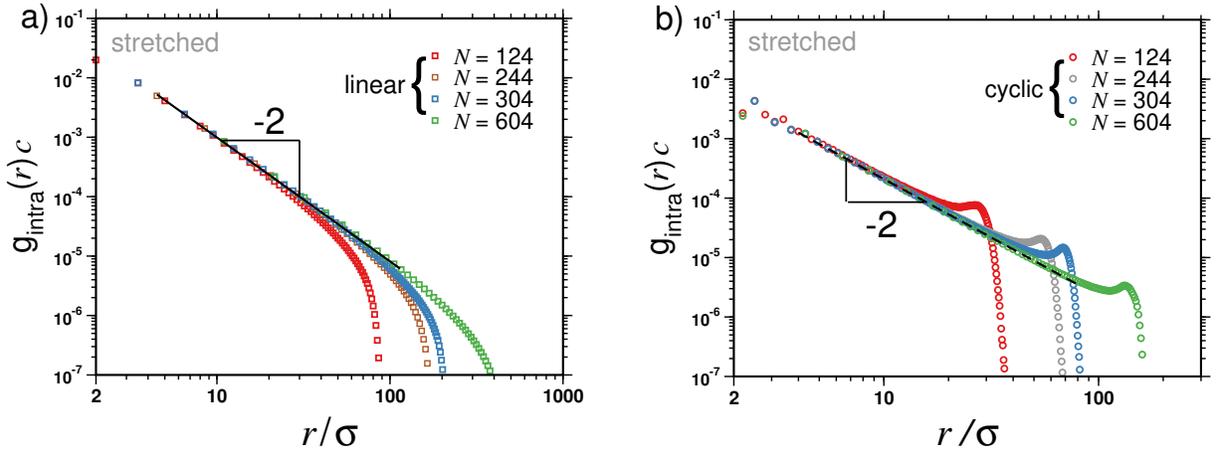

\includegraphics[scale=0.29]{gintra.eps}
\hspace*{.5cm}
\includegraphics[scale=0.29]{gintra_ring.eps}
\caption{\label{fig:gintra}
The intra-chain monomer-monomer correlation functions $g_{\text{intra}}(r)$ for: a) linear (squares) and b) cyclic (circles) polyeletrolytes displayed for
different degree of polymerization $N$
as indicated. All pair correlations functions were rescaled by the overall monomer concentration $c$.  The lines represent theoretical scaling laws predicted for linear (Eq.~(\ref{gL})) and for cyclic (Eq.~(\ref{gC})) architectures. All the data are displayed for fixed Bjerrum lenght $l_B/\sigma =1$.
}
\end{figure}

In Fig.~\ref{fig:gintra}, a comparison is presented between the intra-chain pair correlation functions obtained through analytical calculations (lines) and MD data (symbols) for linear and cyclic architectures with various values of $N$.  In both cases, the functions $g_{\rm intra}(r)$ exhibit a scaling law and decay as $\propto r^{-2}$ for distances $r\ll R_e$ in the reference to linear PEs (solid line), and for distances $r\ll R_s$ in the case of cyclic PEs (dashed line). As noted above the average monomer density $\rho(z)$ for nonuniformly stretched linear PEs  varies logarithmically along the elongation axis (cf.~Eq.~(\ref{rhoz})). The summation in $g_{\rm intra}(r)$ (as seen in Eq.~(\ref{gintra_def})) encompasses all possible monomer pairs. Consequently, this leads to an additional averaging of the monomer density $\rho(z)$ along the chain's stretching direction, where both ends and middle sections contribute to $g^{\rm L}_{\rm intra}(r)$ (cf.~Eq.~(\ref{gL})). As a result of this averaging of the logarithmic function, $g^{\rm L}_{\rm intra}(r)$ closely resembles the correlation function of a uniformly stretched chain, as previously observed for linear PEs in $\theta$ solution.~\cite{liao}.

The primary distinction between the intra-chain functions  for linear and cyclic architectures is the non-monotonic behavior of $g_{\rm intra}^C(r)$, which becomes evident at distances comparable to $R_s$. This distinct property, a hump in $g_{\rm intra}^C(r)$, does not solely arise due to the absence of chain ends, but also originates from the shape of cyclic PEs. These molecules exhibit a slightly oval shape, as evidenced by the eigenvalue ratio $\langle \lambda_1 \rangle / \langle \lambda_2 \rangle \approx 1.2$. This deviation from a perfectly symmetric ring conformation, where $\langle \lambda_1 \rangle/ \langle \lambda_2 \rangle=1$, results in a non-monotonic correlation function as displayed in Fig.~\ref{fig:gintra}b. It is worth noting that when the architecture is arranged regularly on a planar circle, the corresponding intra-chain correlation function lacks a pronounced hump and resembles the linear correlation function $g^{\rm L}_{\rm intra}(r)$ shown in Fig.~\ref{fig:gintra}a.


\subsubsection{Structure factors of linear and cyclic PEs}

\begin{figure}[!h]
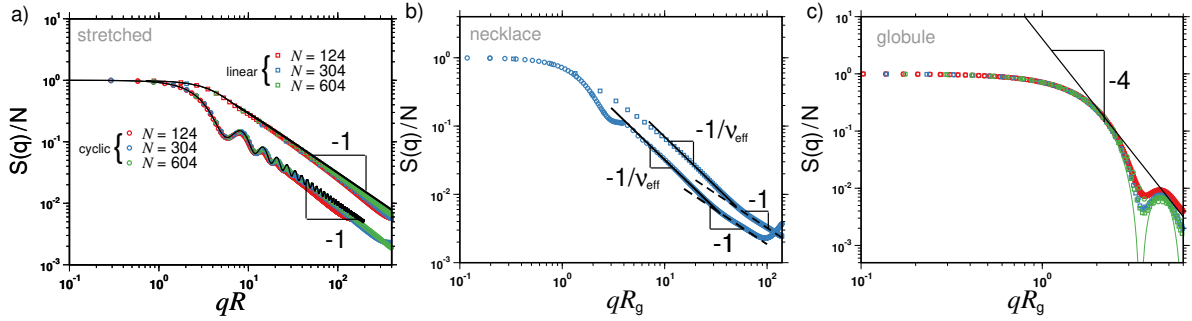

\includegraphics[scale=0.2]{Sq_ring_v2.eps}
\includegraphics[scale=0.2]{Sq_ring_mediumLB.eps}
\includegraphics[scale=0.2]{Sq_ring_largeLB.eps}
\caption{\label{fig:sf}
Form factors $S(q)$ of linear PEs (squares) and cyclic PEs (circles) for different conformational regimes observed at varying values of the Bjerrum length $l_B$: a) stretched ($l_B/\sigma=1$), b) necklace  ($l_B/\sigma=7.5$) and c) globule ($l_B/\sigma=27$). In panel a), the abscissa is rescaled by the root mean-square end-to-end distance $R_e$ for linear chains and by the root mean-square spanning distance $R_s$ for cyclic chains. In panels b) and c), the abscissa is rescaled by the root mean-square radius of gyration $R_g$ for both molecular architectures. The lines represent theoretical scaling laws $S(q)\propto q^{-\alpha}$, where $\alpha$ is the scaling exponent (see text for details).  The data are displayed for various degrees of polymerization $N$ as indicated.}
\end{figure}

To allow direct comparison between experiments and our results, we calculate the structure factors for linear and cyclic topologies. The form factors, $S(q)$, for all three electrostatic coupling regimes and various degrees of polymerization  are shown in Fig.~\ref{fig:sf}. In particular, at low and intermediate Bjerrum lengths,  we observe a striking difference between linear and cyclic form factors. For linear PEs (squares) in the stretched regime, as $q$ increases $S(q)$ exhibits a power-law decay with a slope of $-1$, which is typical of rod-like molecules (Fig.~\ref{fig:sf}a). This indicates that the chain is stretched at all length scales up to a blob size $D_e$, which contains few monomers. 
The simulation data for linear PE chains were also accurately aligned with the theoretical prediction of the form factor $S_{\rm rod}(q)$ for rod-like molecules outlined in Eq.~(\ref{Srod}) using a parameter-free fitting approach.

In the stretched regime, cyclic PEs (circles) also exhibit a power-law decay in $S(q)$ at high values of $q$. However, their behavior in the intermediate range of $q$  is different, as shown in Fig.~\ref{fig:sf}a. Specifically, cyclic chains display sinusoidal undulations superimposed upon the underlying power-law decay, characterized by a slope of $-1$. 
These undulations were fitted to the theoretical prediction for the form factor, $S_{\rm rigid\,ring}(q)$, of a rigid ring (see Eq.~(\ref{Sring})) without involving any adjustable parameters.  It is worth noting that similar undulation patterns were observed in $S(q)$ of semiflexible and rigid neutral cyclic polymers in recent theoretical and experimental studies \cite{yamakawa,sato1,sato2}.

The undulations observed at the weak coupling regimes become weaker as the PE chains enter their necklace regimes at the intermediate values of the Bjerrum length (i.e., $7\lesssim l_B/\sigma\lesssim 9$ (Fig.~\ref{fig:sf}b). As the electrostatic coupling increases, chains become more flexible due to condensation of counterions on their backbones. At large $q$ values, the chain remains stiff up to sections composed of several monomers, which results in a slope of $-1$. At intermediate $q$ values, chains become coiled which is indicated by a transition to a slope of 1/$\nu_{\rm eff}$, where $\nu_{\rm eff} \approx 0.6$ at $l_B/\sigma=7.5$.  

In the strong coupling regimes, for which $l_B/\sigma \gg 1 $, the difference between linear and cyclic chains vanishes since both chains exist in highly collapsed conformation (Fig.~\ref{fig:sf}c). At the large wave-vector limits, we observe a strong decay with an  exponent $-4$. This strong decay reflects the presence of well-defined interfaces and boundaries within the sample~\cite{porod}. The form factors for both topologies 
were accurately aligned through parameterless fitting
  to the theoretical structure factor, $S_{\rm sphere}(q)$, of a homogeneous sphere (cf.~Eq.~(\ref{Ssphere})).


\section{Conclusions}

In this study, we apply scaling arguments that were originally formulated for linear PE chains to construct a theoretical framework. This framework allows us to estimate equilibrium chain sizes and monomer distributions for cyclic PE chains, which we subsequently compare to the outcomes of our coarse-grained molecular dynamics simulations. Our analysis demonstrates that, to a certain extent, charged cyclic polymers exhibit conformational properties that are similar to those of linear PE chains across a range of electrostatic coupling regimes, encompassing weak, intermediate, and strong electrostatic interactions (corresponding to small, intermediate, and large Bjerrum lengths). As a result, in these three regimes, we observe stretched, pearl-necklace, and globular-like conformations for both types of topologies. 
Nevertheless, the resemblance between linear and cyclic architectures diminishes when examining their shapes, the chain length dependence of their characteristic sizes and structure factors, particularly in the weak and intermediate coupling regimes.
Notably, two metrics that describe the shape of PE chains, namely asphericity and prolatness, exhibit substantial differences between cyclic and linear chains. Specifically, they display a non-monotonic dependence upon increasing Bjerrum length for cyclic chains, whereas for linear chains, they exhibit a consistent, monotonous decrease as the electrostatic strength is increased. 
In the stretched regime, the shape of cyclic PEs resembles a flat, two-dimensional ring-like conformation with a substantial surface area. In contrast, linear PEs adopt nearly rod-like, one-dimensional conformations.
 Our analysis also reveals that cyclic PE chains are much more compact in the stretched and necklace regimes as compared to linear PEs. 
Furthermore, the logarithmic correction term present in the scaling law describing the dependence of the average size of linear PEs on the degree of polymerization is absent in the case of cyclic chains due to their circular geometry. This distinction could potentially contribute to the unique conformational behaviors observed in cyclic chains. Specifically, the cyclic topology gives rise to undulations resulting from structures forming over several monomers or more, as evidenced by the structure factor analysis.  In conclusion, our study demonstrates that long-range electrostatic interactions can exert a significant influence on the conformational behavior of cyclic chains, highlighting the need for further {  theoretical and} experimental investigations in the future. { Our forthcoming research will focus on conducting a thorough examination of the effects of salt and ion multivalency on the conformations of cyclic polyelectrolytes. Additionally, we intend to investigate how the cyclic topology influences the dynamics of polyelectrolytes.}

\appendix
\section{Appendix}
\renewcommand{\theequation}{A\arabic{equation}}
\setcounter{equation}{0}
The form factor of a rigid rod of length $R_e$ is\cite{neu}
    \begin{equation}
S_{\rm rod}(q) = \frac 2{qR_e} \left(\int_0^{qR_e} \frac{\sin{x}}{x}dx - 2\frac{\sin^2{[qR_e/2]}}{qR_e}\right).
    \label{Srod}
\end{equation}
The form factor of a rigid ring of diameter $R_s$ reads\cite{oster,huber}
\begin{equation}
S_{\rm rigid\,ring}(q) = \int_{0}^{\pi/2} d\varphi [J_0(R_sq\sin{\varphi})]^2 \sin{\varphi},
    \label{Sring}
\end{equation}
where $J_0(x)$ is the Bessel function of the first kind with order 0.  \\
The form factor of a sphere with radius of gyration $R_g$ is given by\cite{ray}
\begin{equation}
S_{\rm sphere}(q) = \left[ \frac{3}{(\sqrt{5/3}qR_g)^3}( \sin{[\sqrt{5/3}qR_g]} - \sqrt{5/3}qR_g\cos{[\sqrt{5/3}qR_g}]    )   \right]^2.
\label{Ssphere}
\end{equation}

\begin{acknowledgement}
This work has been supported by the National Science Center, Poland (Grant Sonata Bis No.~2018/30/E/ST3/00428). We thank M.~Lang for fruitful discussions.  We are also grateful to PL-Grid Infrastructure for a generous grant of computing time.
\end{acknowledgement}

\bibliography{apssamp.bib}

\newpage
\begin{center}
\Large{Table of Contents Graphic}
\end{center}
\vspace{1cm}

\begin{figure}[t]
\centering
\includegraphics[width=8.25cm,height=4.45cm]{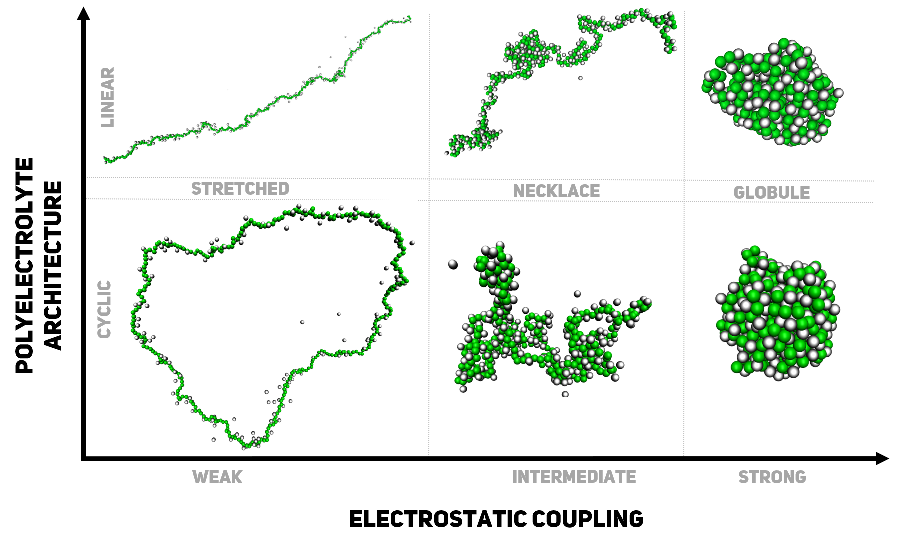}
\end{figure}

\vspace{1cm}

\end{document}